\documentstyle[11pt,paspconf]{article}
\begin{document}

\title{Numerical Data-Processing Simulations
of Microarcsecond Classical and Relativistic Effects in Space Astrometry}

\author{Sergei M. Kopeikin}
\affil{Department of Physics \& Astronomy, University of Missouri-Columbia, 223 Physics Building, Columbia, MO65211, USA}
\author{N.V. Shuygina, M.V. Vasilyev, E.I. Yagudina}
\affil{Institute of Applied Astronomy, RAS, 10, Kutuzov quay, 191187, St.Petersburg, Russia}
\author{L.I. Yagudin}
\affil{Pulkovo Observatory, St.Petersburg, Russia}

\begin{abstract}
\scriptsize The accuracy of astrometric observations conducted via a space-borne 
optical interferometer orbiting the Earth is expected to approach a few
microarcseconds. Data processing of such extremely high-precision 
measurements requires access to a rigorous relativistic model of light ray propagation developed in the framework of General Relativity. The data-processing of the space interferometric observations must rely upon the theory of general-relativistic transformations between the spacecraft,
geocentric, and solar barycentric reference systems allowing unique and unambiguous interpretation of the stellar aberration and parallax effects. 
On the other hand, the algorithm must also include physically adequate treatment of the relativistic effect of light deflection caused by the spherically-symmetric (monopole-dependent) part of the gravitational field of the Sun and planets as well as the quadrupole- and spin-dependent counterparts of it. In some particular cases the gravitomagnetic field induced by the translational motion of the Sun and planets should be also taken into account for unambigious prediction of the light-ray deflection angle. In the present paper we describe the corresponding software program for taking into account all classical (proper motion, parallax, etc.) and relativistic (aberration, deflection of light) effects up to the microarcsecond threshold and demonstrate, using numerical simulations, how observations of stars and/or quasars conducted on board a space optical interferometer orbiting the Earth can be processed and disentangled. For doing numerical simulations the spacecraft orbital parameters and the telescope optical-system-characteristics
have been taken to be similar to those in the HIPPARCOS mission. The performed numerical data analysis verifies that the relativistic algorithm chosen for data processing is convergent and can be used in practice for determining astronomical coordinates and proper motions of stars (quasars) with the required microarcsecond precision.
Results shown in the paper have been obtained with the rather small number of stars (a few thousand). Simulations which are based on a much larger number of stars taken, e.g., from the Guide Star Catalogue used for modelling original observations are to give more complete information about potential abilities of the space astrometric missions.
\end{abstract}
\small
\section{Introduction}
Current technological achievements have greatly improved the accuracy of optical observations. Nowadays an optical interferometer based on board a spacecraft orbiting the Earth is considered to be the most promising tool for fundamental astrometry and astrophysics. Microarcsecond (and, probably, sub-microarcsecond) accuracy in measuring angular distances between pairs of point-like celestial objects (stars, quasars, etc.) in the sky could become feasible. In practice, the accuracy of the planned space astrometric missions is anticipated to reach 50 $\mu$as [\cite{1}] through 4 $\mu$as [\cite{2}] to 1 $\mu$as [\cite{3}]. It is commonly recognized [\cite{4}]-[\cite{5a}] that data processing of optical observations attaining the above-specified accuracy can not be tackled in the framework of the classical approach based on the Newtonian concepts, which must be abandoned and replaced with one based on general relativity. Indeed, the post-post-Newtonian deflection of light caused by the Sun as well as post-Newtonian deflections of the light ray associated with
the oblateness of the Sun and major planets, and their translational and
rotational motions can achieve values of order of 1-10 $\mu$as [\cite{5}].

    The fact that observations are carried out on board the earth-orbiting satellite requires precise definition of the satellite, geocentric, and solar barycentric reference frames along with the corresponding relativistic transformations between them. Combining relativistic description of the light ray trajectory, written in the solar barycentric reference frame, with the relativistic transformations between the different frames allows to develop  unambiguous models of stellar aberration [\cite{6}], annual and diurnal parallaxes [\cite{7}]-[\cite{8}] of stars. Unfortunately, the complexity of this mathematical model makes it difficult to estimate the real accuracy of measurements of classical and relativistic effects. Furthermore, it is not self-evident that the model in question will make it possible to disentangle billions of measurements of arcs between pairs of different stars in order to get their astronomical positions,  proper motions, and parallaxes. For these reasons, one needs to study these and related questions using the Monte-Carlo numerical simulations of the space interferometric optical data-processing before the mission is launched.

For the numerical study of the problem under consideration the software called Ephemerides
Research in Astronomy (ERA) and developed in the Institute of Applied Astronomy [\cite{9}] has been applied. This software was designed in order to provide a universal method of data processing of different types of astronomical observations --- radio, optical, X-ray, etc. --- in ephemeris astronomy. 
Unfortunately, the classical model of the data reduction of optical astrometric observations adopted in ERA did not take into account all relativistic
effects properly. Therefore, the goal of the present work is to re-design the model of the data reduction procedure used in ERA on the basis of a more advanced relativistic model. In this paper we, first, describe the main modifications used in ERA extending its applicability to reach the 
microarcsecond threshold and, second, apply the modified ERA for numerical study of processing of optical data obtained with microarcsecond accuracy.  

\section{Relativistic Model for Data Processing in Microarcsecond Space Astrometry}

In order to avoid misinterpretation of observed astrometric data made in space at the microarcsecond level of accuracy a general relativistic description of light propagation from the source of light to observer and that of reference frames in the solar system are required. The most complete and successful theoretical study of this problem was undertaken in a series of our publications [\cite{5}]-{\cite{7}], [\cite{10}] triggered originally by the launch of HIPPARCOS [\cite{11}] space-mission and the work on the POINTS [\cite{12}] project. 
The approach under discussion is based on the post-Newtonian theory of astronomical reference systems and the mathematical technique of matched asymptotic expansions [\cite{13}]. Three basic coordinate charts are constructed by solving the Einstein equations iteratively in the slow-motion and weak-field approximation --- the barycentric reference system (BRS) of the solar system, the geocentric reference system (GRS), and the satellite reference system (SRS). BRS covers the whole space-time and is used to describe light propagation from the observed celestial object to observer as well as to study the motion of bodies around the Sun. GRS covers a space domain in the vicinity of the world line of the earth's center of mass and is used to trace the motion of the earth-orbiting satellite carrying with the the space optical observatory. SRS represents a local coordinate system of observer (optical interferometer) placed at its origin.

The algorithm of reduction of space astrometric observations performed on board earth-orbiting satellite would involve relativistic transformation of star's position observed in SRS to the barycentric unit vector of the star defined by the extrapolation of the light-ray trajectory from the star to observer to the past-null infinity of the asymptotically-flat, Minkowskian space-time where the inertial reference frame is formally defined\footnote{The advanced post-Minkowskian theory of light propagation including explanation of the relativistic terminology, used for the description of microarcsecond astrometric observations, has been recently constructed in [\cite{7}] (see also the paper by Kopeikin \& Gwinn in this proceedings).}. More specifically, the data reduction procedure includes the following steps [\cite{6}]: 
\begin{enumerate}
\item Components of the observed unit vector of a star $\vec{s}$ defined at the origin of the SRS are transformed to those of the unit vector $\vec{p}$ defined in the BRS using the relativistic transformations from SRS to GRS and, then, from GRS to BRS. This two-step transformation eliminates stellar aberration caused by the motion of satellite about the Earth and the Earth about the Sun.
\item The positional unit vector $\vec{p}$ is attached to the point of observation and still includes relativistic deflection of light due to the gravitational field of the solar system. It is eliminated by means of mapping $\vec{p}$ to the asympotically flat space-time using solution of equations of light propagation. This procedure transforms the positional vector $\vec{p}$ to the vector $\vec{k}$ lying in the flat space-time and attached to the point of observation. Microarcsecond accuracy of observations necessitates at this step to account for both the post-Newtonian (1PN) and the post-post-Newtonian (2PN) perturbations in the star's position. More specifically, in the 1PN approximation deflections of light due to the spherically-symmetric part of the gravitational field of the Sun, Moon, Earth, and major planets must be subtracted along with the deflections caused by the orbital velocity of Jupiter and quadrupole component of the field of the major planets and the Sun.  The 2PN light deflection is noticeable only for the gravitational field of the Sun.
\item The last step of the reduction procedure involves taking into account the parallactic dispacement and proper motion of the star. This is done by making an ordinary Taylor expansion of the vector $\vec{k}$ with respect to the parallax $\pi$ and time $t-t_0$ increments where $t$ is the time of observation and $t_0$ is the initial epoch of observations. In this way we obtain the astrometric position, proper motion, and parallax of the star which can be further used for the catalogue's composition.    
\end{enumerate}
Let us now describe how the model works in the analysis of numerically simulated observations.
\section{Numerical Simulation of Astrometric Observations and Their Processing}

The modified program package ERA [\cite{9}] was used as a main tool for
the numerical simulation of observed positions of celestial objects. The most significant modifications have been done in that part of the program dealing with
the calculation of the relativistic light-ray deflections and the SRS-GRS-BRS space-time transformations. The new program has been carefully tested by calculating residuals in star positions with respect to the previous version of ERA. The residuals are in excellent agreement with the theoretically predicted
deviations between the old and new versions of ERA.

For the simulation of the space astrometric experiment a list of 13000 stars having random proper motions and radial velocities and uniformly distributed all over the celestial sphere has been generated and treated as the input stellar catalog in which positions, proper motions and radial velocities of the stars were considered as precisely known\footnote{We did not include parallaxes of stars in the catalogue. They are assumed to be taken into account in the future, more enhanced study.}. It was also assumed that all stars in the catalog have brightnesses that are sufficient in order to reach a microarcsecond observational precision. A fictitious observer was placed at the geostationary satellite carrying three telescopes with optical axes lying in one plane (the satellite "equatorial" plane) and separated at angles $54^\circ$, $78^\circ$ and $132^\circ$. Each telescope was assumed to have a strip-like coherent field of observation $1.6^\circ$ wide. The satellite is assumed to be rotating around the spin axis, which is perpendicular to the "ecuatorial" plane, according to a scanning law tha is similar to that used in the HIPPARCOS mission\footnote{The spin axis
makes a constant angle of $43^{\circ}$ with respect to the direction to the Sun and precesses around this direction at the rate of 6.4 rev/yr. The satellite's period of rotation around the spin axis was assumed to be 128 minutes.}

The fake-observations generating code applies the procedure described in section 2 in the reverse order to simulate positions of stars as they would be visible from the satellite. The random error of a single "observation" was taken to be 1$\mu$as.  Then, the program starts to rotate the satellite at the prescribed angular velocity and "observations" commence at some epoch $t_0$ such that the telescopes sweep over the simulated positions of stars step by step. The program calculates the telescope's directions with respect to SRS and looks through the input catalogue reduced to the SRS frame choosing pairs of stars which are available for observations simultaneously by two of three telescopes. Each "observed" pair of stars gives one conditional equation for determining parameters of the stars and magnitude of measurable classic and relativistic effects. Due to the limited computer resources the simulation of observations have been done at the span of one year with the step of 0.1 day. This gave us approximately 3000 observed arcs between pairs of stars on the sky. For the subsequent parameter estimation we selected only those stars in the arcs which had more than 5 "observations". The number of such stars reached 1000. Their "observed" positions were processed by the least-squares-method in accordance with the model of section 2 and the post-fit astrometric positions of stars were compared with those from the initial input catalogue. The next section presents the results of the  applied procedure along with a brief discussion. 

\section{Results and Discussion}

The basic results of our numerical study of the problem of data processing of observations done with a microarcsecond accuracy on board an artificial satellite orbiting the Earth are as follows: 
\begin{itemize}
\item Even a small number of observations with the precision of a single measurement approaching 1 $\mu$as will make it possible to improve presently-known astrometric positions, proper motions, and parallaxes of stars\footnote{In our numerical simulations the post-fit accuracy, e.g., astrometric positions of stars is slightly better than 40 $\mu$as.}. Relativistic effects then also be measured with improved accuracy.
\item There can exist strong correlations between classic astrometric parameters of stars and those characterizing relativistic effects. The correlations get weaker when more observations are included and/or distribution of stars all over the sky tends to be more uniform. In addition, it is worth noting that the number of relativistic parameters depends on the relativistic model one uses for data processing. More straightforward models giving a smaller number of relativistic parameters will be more preferable for space astrometry.
\item The number of observations and the observed distribution of stars may crucially restrict the feasibility of microarcsecond space astrometry. More powerful computer codes and/or much more realistic numerical simulations including billions of stars are required to answer the question whether microarcsecond threshold in measuring classic and relativistic effects can be reached (see, e.g., [\cite{15}]).  
\end{itemize}
\acknowledgements
We thank Bahram Mashhoon for a critical reading of the manuscript and useful comments. We are grateful to Victor Brumberg and Sergei Klioner for valuable discussions.  

\end{document}